\documentclass[ag]{copernicus}

\usepackage{mathptmx}

\usepackage{graphicx} 

\usepackage{amsmath}
\usepackage{amsfonts}
\usepackage{amssymb}
\usepackage{amsbsy}
\usepackage[figuresright]{rotating}
\usepackage{multicol}
\usepackage{array}
\usepackage{supertabular}

\usepackage{color}

\begin{document}

\title{{Magnetopause displacements:  The possible role of dust}}

\author[1,2]{{R. A. Treumann}
}
\author[3]{{W. Baumjohann}}

\affil[1]{Department of Geophysics and Environmental Sciences, Munich University, Munich, Germany}
\affil[2]{Department of Physics and Astronomy, Dartmouth College, Hanover NH 03755, USA}
\affil[3]{Space Research Institute, Austrian Academy of Sciences, Graz, Austria}

\runningtitle{Magnetopause displacement}

\runningauthor{R. A. Treumann and W. Baumjohann}

\correspondence{R. A.Treumann\\ (rudolf.treumann@geophysik.uni-muenchen.de)}

\received{ }
\revised{ }
\accepted{ }
\published{ }


\firstpage{1}

\maketitle

\begin{abstract}
Large compressions of the magnetopause are proposed to occasionally result from temporary encounters of the magnetosphere with dust streams in interplanetary space. Such streams may have their origin in cometary dust tails or asteroids which cross the inner heliosphere or in meteoroids in Earth's vicinity. Dust ejected from such objects when embedding the magnetosphere for their limited transition time should cause substantial global deformations of the magnetopause/magnetosphere due to the very large dust grain mass and momentum which compensates for the low dust density when contributing to the upstream pressure variation.  

\end{abstract}

\vspace{3mm}\noindent
The magnetopause is almost constantly located with its nose at a nominal average geocentric solar wind stagnation point distance $10\ \mathrm{R}_\mathrm{E}\lesssim r_\mathit{mp}< 11\ \mathrm{R}_\mathrm{E}$. Being a free boundary between plasmas of different properties \citep{spreiter1966}, it is a highly dynamical semi-transparent surface of separation which, quite naturally, is oscillating around its nominal distance. Such oscillations may be quasi-periodic being caused by fluctuations in the solar wind and magnetosheath pressure and magnetic field, turbulence and surface waves. The latter being caused by streaming instabilities like the Kelvin-Helmholtz \citep{chandra1957,pu1983,hasegawa2004} or Kruskal-Schwarzschild \citep[for a recent discussion see e.g.][]{plaschke2011} instabilities. Variations in reconnection efficiency can also cause small displacements of the magnetopause and, to some limited extent, internal magnetospheric processes may sometimes be involved. Not all of these processes have been sufficiently understood so far. 
\begin{figure*}[t!]
\centerline{{\includegraphics[width=0.8\textwidth,clip=]{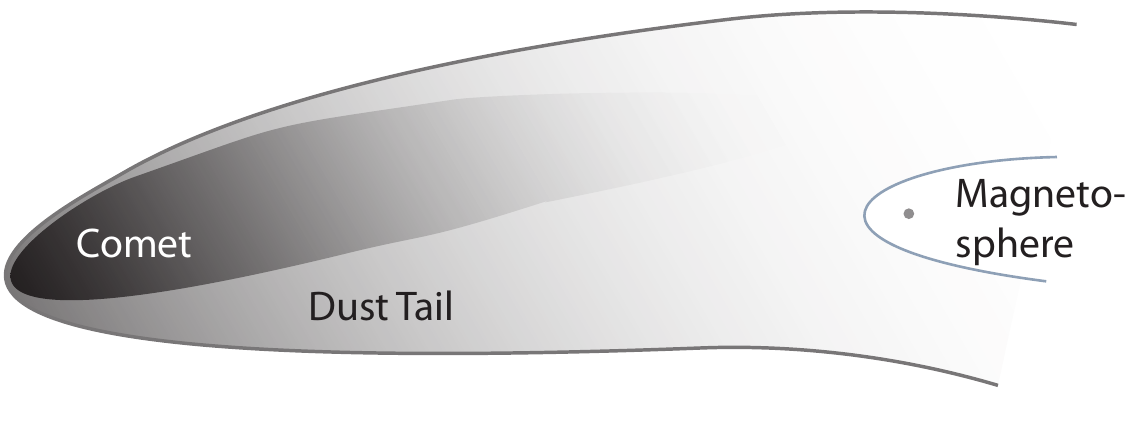}
}}\vspace{-3mm}
\caption[ ]
{\footnotesize {Schematic of the magnetosphere embedded into a wide cometary dust tail.}}
\label{fig-comtail}
\end{figure*}

Oscillations may also occur non-periodically, however.  Occasional very large compressions of the magnetosphere accompanied by displacements of the magnetopause close to the Earth are difficult to explain by any of these processes -- with the exception of violent Coronal Mass Ejection (CME) events --, in particular by the mere increase of solar wind pressure or/and the enhancement of reconnection at the dayside magnetopause. The stagnation point distance depends only very weakly on the solar wind pressure, and the reconnection efficiency barely exceeds a factor of $\sim0.1$, i.e. $\sim10\% $ in the magnetic field strength, not enough for causing a large displacement of the magnetopause stagnation point. In some cases strong compressions of the magnetosphere have indeed been unambiguously related to ordinary though extreme solar wind pressure increases caused by the mentioned Coronal Mass Ejections \citep[cf., e.g.,][and references therein]{shue1998}. Such events are accompanied by measurable strong variations in the solar wind parameters, density, velocity, and temperature which the CME plasma caused in the solar wind. They can thus easily be monitored. 

In the absence of a CME large magnetopause excursions are more difficult to understand.
Surface wave excitation does barely cause the required very large amplitudes which are needed in order to displace the magnetopause stagnation point region substantially closer to Earth; they moreover are of short tangential wavelengths such that the displacement would be observed not globally but locally only.  Such occasional local intrusions have tentatively been attributed to the presence of PTEs,  so-called magnetosheath Plasma Transfer Events \citep[also known as IPEs, Impulsive Penetration Events,  see][]{lemaire1977,heikkila1982,owen1991,sibeck1989,sibeck1990,brenning2005}, or -- in the case of localised  expansions -- the presence of so-called HFAs, Hot Flow Anomalies, in the solar wind \citep{schwartz1988,schwartz1995,sibeck1999,korotova2011}, respectively. But rare large-scale  compressions of the magnetosphere down to stagnation point distances of $r_\mathit{mp} \sim8\ \mathrm{R}_\mathrm{E}$ or even $r_\mathit{mp}\sim 6\ \mathrm{R}_\mathrm{E}$ unrelated to either kind of such obvious events remained unexplained.

In the present Note we explore the possibility whether occasional clouds of interplanetary dust could as well take responsibility for large if not to say extreme compressions. These would not be accompanied by any detectable solar wind plasma pressure increases thus resisting an interpretation by ordinary plasma effects. That dust exists in interplanetary space in various forms is a well known fact \citep[cf., eg.,][and references therein]{grun2011,kempf2003,kruger2009}. These forms reach from interstellar dust to planetary dust, dust from asteroids and cometary dust being of variable sizes and masses. Usually dust is very dilute and does not play any dynamical role in the plasma. However, during passages of Comets \citep{agarwal2010,fulle2010,sykes1986,sykes2004} or smaller Asteroids closer to the Sun as well as due to other reasons like meteoroids \citep{staubach1997} it may occasionally happen that the number density of dust particles is temporarily and locally  enhanced in the solar wind or, to be more precise, in a solar wind flux tube that would be connected to the dust source, Asteroid, Comet or meteoroid. In such a case, when contacting the magnetosphere one may expect that the presence of the dust makes a difference.

Dust particles are extraordinarily heavy with masses $m_d=A_dm_p$ large multiples $A_d\gg1$ of the proton mass $m_p$ (in other words, the atomic number of the dust grain). In addition, exposed to solar ultraviolet light illumination and collisions with solar wind particles, they necessarily become charged. Since collision frequencies are low their charging $q_d=eQ_d$ is mainly due to photo-ionisation by solar UV and will hence be positive and large: $Q_d\gg1$. Such particles represent very massive ions of $\gtrsim$\, nm radius but can be also larger, up to mm size in the solar wind where they are, however, much less abundant. A most appropriate size is apparently $\sim 5 \mu$m here. Reviews of the physical properties of such dust particles are abound \citep[cf., e.g.,][and references therein]{ingham1961,perrin1989,shukla2002}. 

Densities of dust in the solar wind are small, even though it is known that the Earth accretes several $10^4$ tons of dust per year \citep[cf., e.g.,][]{love1995}. A steady dilute dust distribution in interplanetary space will not become remarkable in its effect on the magnetopause because such dust behaves like a rare single particle component. However, the collective effect of dust may become appreciable when the dust flux increases occasionally in cases like those mentioned above. For example, Figure \ref{fig-comtail} shows schematically the extreme case of a passage of the magnetosphere across a cometary dust tail. More realistic is, however, that the Magnetosphere is briefly hit by a dust cloud which is expelled from a cometary tail or from an asteroid passing across the inner heliosphere.

What concerns the effects and properties of dust we are interested only in the contribution of dust to the solar wind dynamical pressure as this is the only place where a direct affect is expected to be seen on the location of the magnetopause as follows \citep[cf., e.g.,][]{treumann1996} from the well-known basic expression for the stagnation point distance
\begin{equation}\label{eq-1}
R_\mathit{mp}=\left(\frac{KB_0^2}{2\mu_0N_\mathit{sw}M_\mathit{sw}V_\mathit{sw}^2}\right)^\frac{1}{6}
\end{equation}
of the magnetopause. $K=\mathrm{O}(1)$ is (the usually used) numerical factor accounting for the deviation of the geomagnetic field at the magnetopause distance from dipolar geometry, $B_0$ Earth's surface magnetic field, $R_\mathit{mp}=r_\mathit{mp}/\mathrm{R}_\mathrm{E}$ the stagnation point distance measured in Earth radii, and $(N,M,V)_\mathit{sw}$ are the effective solar wind density, mass, and velocity including all solar wind particle components. These latter quantities are defined in the centre-of-mass system of the solar wind and thus need to be determined before proceeding.

We first observe that the densities of the various solar wind components are not independent. They are related through the charge neutrality condition which, due to the large charging of the dust particles, becomes non-trivial. Assuming solar wind protons only we have
\begin{equation}\label{eq-2}
N_e=N_p+Q_dN_d
\end{equation}
where $N_e$ is the original solar wind density $N\equiv N_p$ plus the number of electrons produced by photo-ionisation and added to the electron density. This number is given by the second term on the right. Even though the dust density $N_d$ is low, the large charging $Q_d\gg 1$ of the dust particles under stationary conditions implies a non-negligble effect on the density of mobile electrons. In the extreme case $Q_dN_d\sim N_p$ ions will become a minority of reduced importance locally and the quasi-neutrality condition is replaced by charge balance between electrons and dust. This may have many other effects on the plasma as, for instance, affect reconnection at the magnetopause by the presence of ``very heavy ions" completely determining the scales. Such questions will not be investigated here. 

What concerns the dust velocity, we observe that each dust particle behaves like a pick-up particle when injected at low speed into the solar wind and being charged \citep[cf., e.g.,][for the example of lunar dust particles]{mall1998}. The charging proceeds on a very fast time scale which is almost instantaneous because of the very high UV-photon density near Earth's orbit. Hence, any dust particle immediately experiences the solar wind Lorentz force and within its gyration time $\omega_{cd}^{-1}=m_d/q_dB_\mathit{sw}=(A_d/Q_d)\omega_{cp}^{-1}$ couples to the solar wind and becomes accelerated into the flow. Since both $A_d$ and $Q_d$ are large numbers this time is not too different from the gyration time $\omega_{cp}^{-1}\approx$ few seconds of a solar wind proton even if of the order of minutes or an hour.  In this process the dust particle is accelerated (i.e. heated in the perpendicular direction) to roughly 4 times the (perpendicular) energy, implying that the dust particle because of its large mass $m_d\gg m_p$ still remains cold. On the other hand, its momentum in the solar wind flow 
\begin{equation}
p_d=m_dV_\mathit{sw}\approx A_dp_p
\end{equation}
is much larger than that of the protons $p_pm_pV_\mathit{sw}$, and the total dust momentum density  becomes
\begin{equation}
N_dp_d=A_d\alpha_dp_p, \qquad\mathrm{with}\qquad \alpha_d=\frac{N_d}{N_p}
\end{equation}
the concentration of dust particles. (Remember that $N_p$ plays the role of the dust-free solar wind density.) One may thus expect that even a dilute dust component in the solar wind will have some effect on the pressure equilibrium at the magnetopause.

We need to go to the centre-of-mass frame. Here the mass density is defined through
$NM_\mathrm{sw}=N_em_{\rm e}+N_pm_{\rm p}+N_dm_{\rm d}$ where  the (average) dust mass $m_{\rm d}$ plays an important role. Calculating the
solar wind mass density we have to account for
charge neutrality Eq.~(\ref{eq-2}).  This yields in the centre-of-mass frame
\begin{equation}
A_dN_\mathit{sw}\approx \left[\left(1+\frac{m_e}{m_p}\right) +A_d\alpha_d\left(1+\frac{m_e}{m_p}\frac{Q_d}{A_d}\right)\right]N_p
\end{equation}
The ratio of electron to proton mass in the first parenthesis can be neglected. This expression can then be used in Eq. (\ref{eq-1}) in determining the stagnation point distance of the magnetopause. What concerns the solar wind velocity appearing in that expression so we recall that the dust particles have coupled to the solar wind within one half dust-gyration time performing a cycloidal orbit and thus $V_\mathit{sw}$ is the same for all particles.

Inserting the above expression into Eq. (\ref{eq-1}) including dust of density $N_d$ then yields for the dust-stagnation point distance
\begin{equation}
R_\mathit{mp}^d\approx R_\mathit{mp}^p\left[1+A_d\alpha_d\left(1+\frac{m_e}{m_p}\frac{Q_d}{A_d}\right)\right]^{-\frac{1}{6}}
\end{equation}
According to this expression, the presence of charged dust in the solar wind will, as has been expected,  decrease the stagnation point distance of the magnetopause if only the second term in the bracket is sufficiently large. This term is basically proportional to the ratio of dust-to-proton number density in the solar wind multiplied by the mass ratio $A_d$. The latter is the number of nucleons (protons and neutrons) in a typical dust grain. 

The ratio $Q_d/A_d\ll1$ in the above expression is typically small. This is clear from the fact that only the surface of the dust particle is exposed to illumination by solar UV such that photo-ionisation is restricted to the uppermost layer of the dust grain and does not include all nucleons in the dust grain which contribute to the mass. This  still holds even though dust grains have rather complicated geometric form. Moreover, an atom at the illuminated surface  of the dust grain may become multiply ionised by sufficiently energetic UV photons. Suggested charge numbers are $10^4<Q_d<10^6$. Hence the second term in the parentheses in the above expression, which is multiplied by the electron-to-proton mass ratio, is small, and the reduction of the stagnation point distance depends mainly on the factor $A_d\alpha_d$. Reduction thus occurs whenever $A_d\alpha_d>1$. 
\begin{figure}[t!]
\centerline{{\includegraphics[width=0.5\textwidth,clip=]{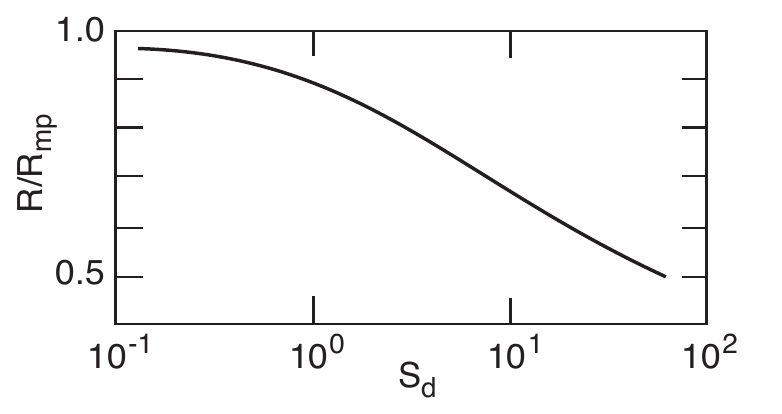}
}}\vspace{-3mm}
\caption[ ]
{\footnotesize {The dependence of the magnetopause stagnation point distance on the function $S_d=1+A_d\alpha_d(1+m_eQ_d/m_pA_d)$. Normalisation is to the nominal stagnation point radius $R_\mathit{mp}$.}}
\label{fig-Sd}
\end{figure}

Dust grain masses are very uncertain due to porousness. Dust grains observed from Comets are of porous material with internal mass density of $n_d\lesssim 2\times 10^{3}$ kg/m$^3$. Assuming that they have radius of few $\mu$m corresponding to a volume of $\sim10^{-17}$ m$^3$, they contain a mass of $M_d\sim 10^{-13}$ kg \citep[e.g.][]{kempf2003} corresponding to $A_d\lesssim 10^{14}$ nucleons, which also justifies neglect of the ratio $Q_d/A_d$. Cometary dust densities far outside the cometary coma in the solar wind have been measured to be of order $10^{-2}\lesssim N_d\lesssim 10^{-1}$ m$^{-3}$. This implies dust concentrations of $10^{-9}\lesssim N_d/N_e\lesssim 10^{-8}$ or
\begin{equation}
A_d\alpha_d \lesssim 10^3 
\end{equation}
in this particular case. Similar numbers can be expected from asteroids and from meteoric showers. For observing an extreme reduction of the stagnation point distance by a factor 2 from $R_\mathit{mp}=10$ to $R_\mathit{mp}=5$, one needs a ratio $S_d\equiv 1+A_d\alpha_d(1+m_eQ_d/m_pA_d)=64$ which is well in the above range. Figure \ref{fig-Sd} shows the dependence of the normalised nose distance $R^d_{mp}/R_{mp}^p$ as function of $S_d$.

These estimates show that whenever the Earth is hit by or crosses a dust stream of, for instance, cometary origin and sufficiently high dust number density,  as shown schematically in Figure \ref{fig-comtail}, then it becomes highly probable that the magnetopause will be much stronger compressed and displaced from its about undisturbed location and shape than by any other mechanism like surface waves or solar wind fluctuations. The compression depends on dust particle mass, photo-ionisation, charging of dust and, of course crucially, on the number density $N_d$ of dust particles. The latter is a function of distance $r$ from the dust source. It decays (approximately) like $N_d\propto ( r/R_\mathit{com})^{-2}$, where $R_\mathit{com}$ is the radius of the cometary ionopause. It also requires that the Earth does indeed enter the dust stream which readily couples to the solar wind and remains for the diffusion time of the dust in the flux tube of the charged dust stream while this stream is flowing downstream with the solar wind. Therefore, the dust stream does not necessarily cross the Earth and its magnetosphere. However, when it happens the effect discussed in this note should occur and could give rise to a very substantial compression and/or deformation of the magnetosphere.

\begin{acknowledgements}
This research was part of an occasional Visiting Scientist Programme in 2006/2007 at ISSI, Bern. RT thankfully recognises the assistance of the ISSI librarians, Andrea Fischer and Irmela Schweizer. He appreciates the encouragement of Andr\'e Balogh, Director at ISSI. 
\end{acknowledgements}

\end{document}